# A Middleware road towards Web (Grid) Services

Zeeshan Ahmed
Department of Intelligent Software Systems and Computer Science,
Blekinge Institute of Technology,
Box 520, S-372 25 Ronneby, Sweden
*zeah*@student.bth.se

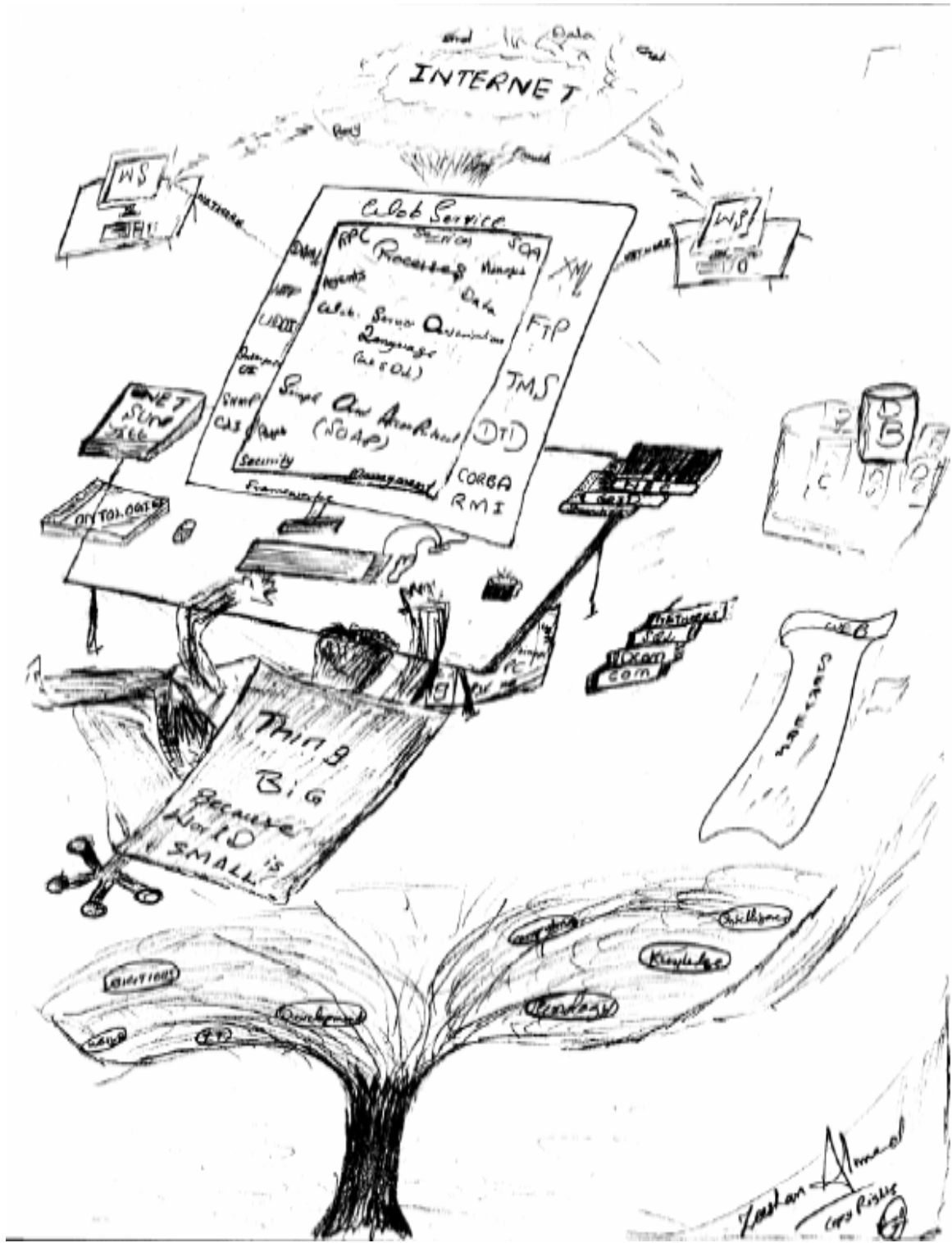

**ABSTRACT:**

Middleware technologies is a very big field, containing a strong already done research as well as the currently running research to confirm already done research's results and the to have some new solution by theoretical as well as the experimental (practical) way. This document has been produced by *Zeeshan Ahmed (Student: Connectivity Software Technologies Blekinge Institute of Technologies)*. This describes the research already done in the field of middleware technologies including Web Services, Grid Computing, Grid Services and Open Grid Service Infrastructure & Architecture. This document concludes with the overview of Web (Grid) Service, Chain of Web (Grid) Services and the necessary security issue.

## 1. INTRODUCTION:

In these days the technologies is growing like a big tree whose every branch is containing a new way towards advancement and every leaf is containing new idea. One of the branches of this tree is Middleware technologies, and this branch is becoming large with more leaves and fruits. Middleware technologies are related to the distributed and online application areas and the main aim of this is to promote the sharing of the resources between the software while keeping them in distributed or online environment. There is a lot of work already has done before and more research is also going on in the different areas of this fields to improve and to invent.

Web service is one of the new sub branch of this field, which is growing more rapidly and the aim of this to promote the sharing of resources between applications in online environment, and by using this approach a big target can be divide into sub target and can be achieved with very less time. Now agents has also been involves in the web service to have an automated online work environment and this approach is also extending from agents to multi agent online systems.

There is another sub branch of the middleware technologies is Grid Computing, and the aim of this branch is to implements the concept of virtualization to heterogeneous entities to provide distributed infrastructure. This branch is moving to some other sub branches, one of them is Open Grid Service to provide ubiquitous access to system with the implementation of Grid Computing and Web (Grid) Service as the combination of both the Grid Computing and the Web Service to have the benefits of both the Grid computing and the web services in one application area. And the major aim of this technology is to improve the sharing of resources in a fast and secure way with in virtual environments by ubiquitous having access.

## 2. MIDDLEWARE TECHNOLOGIES (MDLW TECH)

This is one of the new and hottest topics in the field of the computer science which encompasses of many technologies to connects computer systems and provide a protocol to interact with each other to share the information between them. The term Middleware carries the meaning to mediate between two or more already existing separate software applications to exchange the data.

Middleware technologies reside inside complex, distributed and online application by hiding their self in the form operating system, database and network details, which mostly user don't want to alter just need the support for the application. This is a decentralized approach to the centralized application to minimize the load and make them in heterogeneous. Middleware services help the developers in a way that they need not to implement the modules for their application to perform the transactions, communications details over network, system maintenance, enough secure environments, perseverance from non volatile data, management of requests and event services. The whole online / distributed application's development architecture has been discussed in figure 1.

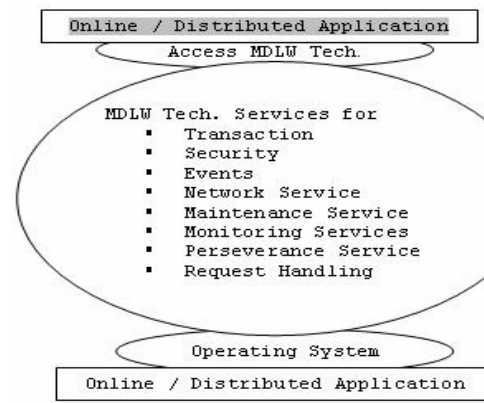

Figure 1
Online / Distriuted Application using Middleware Technologies (MDLW Tech.)

Application just has to access the middleware technologies (MDLW Tech.) and not need to take care of the operating system communication, concurrency services and hardware compatibility issues. By using this approach of software development, developer can save the time and make big reliable application in a very short period of time as compared to the application (expensive & time consuming )which is not using MDLW tech. The main aim is to enhance the reusability of the already made software components.

Most of the MDLW Tech based applications follows the client server architecture to have communication between them, by following the *Remote Procedure Protocol (RPC)* and Messaging by performing Marshalling (convert to network transmission data format) and Demarcating (convert from network transmission data format to actual).

In these days many major companies like Microsoft and Sun are taking so many interests in the middleware technologies because mostly all the advanced planning applications requires more interaction to each other which benefits in providing the good promises to customers as well as taking the fast decisions.

The major middleware technologies are SOAP, Web Services, CORBA, Grid Computing, JAVA, J2EE, SNMP, Jini, EJB and OSGI etc but Web Services and Grid will be discussed in some detail in this paper. These technologies are involved in embedded and enterprise applications, most of all data bases and game applications are focused. Every middle ware technology has its own work flow and based on its won architecture.
More over currently the development in the field of Middleware technologies is going on especially in the fields of Middleware technology development, Grid Oriented Development, Portal Development, E-Science and E-Learning.

## 3. GRID COMPUTING

Grid computing is another middleware technology which implements the concept of virtualization to heterogeneous entities to provide distributed infrastructure by increasing power of computing, storage capabilities & capacity and network speed. The research on this technology is currently also going on. The term *GRID* was initiated in 1990s for advance science and engineering purposes. More over we can say the Grid Computing a kind of distributed system which allows sharing of resources parallel and dynamically at run time but depending upon the availability of those resources. Moreover it also supports the integration of distributed systems, enable interpretability across the borders.

The main aim of Grid technology is to enhance the concept of distribution by enabling different technologies more capable of performing their own tasks in a perfect way by decreasing the time limits, sharing of resources, database solution, and synchronous communication in an abstract way as shown in the Figure Grid.

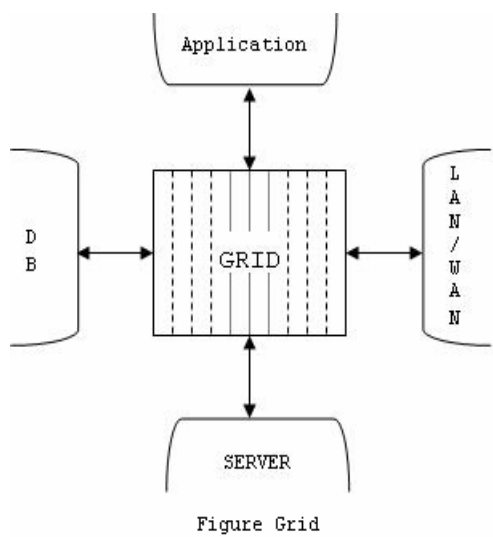

This concept of sharing the data, software and resources to solve a particular problem leads concept of Virtual Organisation (VO) [5]. Virtual Organisation concept is about to create a new domain (administrative) which have their own policies and restrictions.

### 3.1 Grid Architecture

Grid is not massive or huge but it can be the combination of various resources from the various companies. The main components of the Grid are Cluster of computer, Supercomputers, Internet and Job schedules. Gird uses general purpose protocols and interfaces for the communication.

Grid architecture can be divided in to three layers. First layer is low level which contains the Operating system (OS), OS libraries and routines and the Network resources. Second one is the Middle layer which contains the Data, Security Mechanisms (SM), Processes and Languages (Compilers & Interpreters). Third and the last layer is consists of the actual application. As shown in figure Grid Layers.

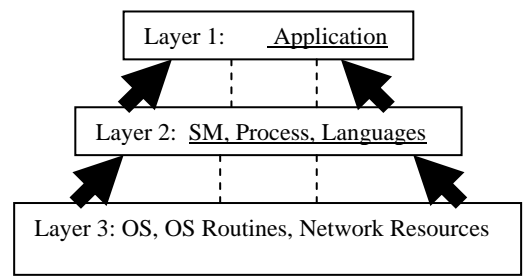

Figure Grid Layers

This architecture provides several benefits to the Grid Computing Applications. At third layer operating system runs and provides all its functionalities based on the its libraries and routines, more over at the third layer the network resources are also initialized and maintained for the communication over the network. Then control moves from third layer to second where actual data is managed to used, languages along with their compilers or interpreters are installed to make new software, actual security mechanisms are also provided and handled at the second level and at the first layer actual application for which the whole layered architecture is made gives the performance.

### 3.2 Grid Computing Security

Grid computing provides the complete mechanism for the security related issues. Methods are provided by the Grid to authenticate the users, to have the secure communication and to have the secure sharing of resources. Cryptography has a good hand behind the Grid's security. Because Grid computing performs cryptographic activities to secure like encryption, decryption, issuing of public keys, private keys and the certificates. Then proper user authentication (single sign on) has been performed on the basis of these. User can only use or share the resources by first passing through the security mechanisms, making his self log in. This kind of cryptography used in the Grid Computing is Asymmetric. That's why this technology considered being secure data transfer technology. [8]

### 3.3 Grid Semantic

This is an approach of Grid Computing, and by using this approach information can be processed by the computer in a standard way. Grid semantic can also viewed as the standard way of sharing various resources and services. This helps in the establishment of connection in the automated way so then resources to be shared in a quick way and to be together to form a virtual organisation. More over it is a kind of grid computing to make the network in a intelligent way, which is aware of all it network components and the network addresses (domain).

## 4. OPEN SERVICE GRID INFRASTRUCTURE (OSGI)

This is the new concept coming from the Grid computing named Open Service Grid Infrastructure. The main aim of the OSGI is to promote virtualization by managing the services and resources for distributed applications based on heterogeneous entities. To have this attitude in the Grid environment, service oriented architecture is required which can be plugged into the Grid architecture. This carries a good scope, because this can serve using as the basis of Web Services (explained in this document in section 7). In OSGI new Port types can be made to enable the standard behaviour.

## 5. OPEN SERVICE GRID ARCHITECTURE (OSGA)

The combination of service oriented architecture and the Grid architecture is formed Open Service Grid Architecture (OSGA) this OSGA should have the ability to manage and share the resources related to network, database and servers. OSGA provides ubiquitous access to system and able to manage the different kinds of work load by reducing the cost and increase the resource utilization. It just likes the blue print of the grid system which defined a good architecture for the Grid Based applications. This provides service based distributed system model and some key features of Grid systems because it is composed of basic behaviours and Grid Services (explained in this document in section 6).

It is platform independent and also capable of creating new service based on already available services, Service to handle the runtime queries, Ability to locate the servers and provide data services to access the data in a consistent and integrated way. More over it should be able to implement the concept of Virtualization and abstraction, capable of performing searching of entities, and Monitor the whole system. OSGA described by Grid Web Service Description Language (GWSDL) which is actually the extension of the WSDL.

OSGA must be secure, as well as the security issue is concerned, the security is required in Authorization in identifying the valid users, integration, accessing resources, cross domain interactions, isolation, delegation and monitoring operations. More over OSGA should able to provide the optimise behaviour at both the client and server ends by improving its quality of services (QoS). This quality of services (QoS) is composed of the commitments between both the service provider and requester and calculating the quality of service provided. Open Service Grid Architecture is also responsible for the execution of several jobs, these jobs requires some kind of management to make available supporting jobs for the execution of the actual complex jobs. This all abilities must be available in distributed environments and across the administrative domains. [7]

## 6. GRID SERVICES

Grid provides services to have ubiquitous access the system, application with the implementation of Grid Computing, Integration of Grid Services, security, reliability, ontologies on Grid, virtual catalo access, logging and session maintenance and configuration management.

The client can access the Grid service by two main instances Grid Service handles (A pointer of network pointing to the instance of Grid Service but it doesn't contain any of required information and then this resolves into the grid service reference) & Grid Service Reference ( A pointer of network but this contains the required information. This is not the permanent pointer).

There is a mechanism to resolve the Grid Service handles to Grid Service Reference which is called Handle Resolver. The grid service created by the request of client for the instance of Grid service for a particular life time period and when that time has been over the automatically the instance has been destroyed (Instance can also explicitly be destroyed). To have a secure interaction between the client's object and the requested service object there should be some abstract communication protocol is required. [6]

These services are the entities of Grid. When we consider a single grid service (entity), then we come to know the grid service is like a web service which has interfaces and the behaviours. More over there are some WSDL conventions are used in the grid services.

## 7. WEB SERVICES

This platform & language independent technology is one of the middleware technologies, based on XML standards for interoperability and some transport protocols to exchange data between the applications. Web Services used to help the people in a way by providing some material to perform some tasks, facilities and efforts to improve their present condition with in possible less time and cost. So if we look in this way so we can say Web Services used by the developers in making transaction, reliable messages, implementing security, pragmatic interface across the networks and a standard way to integrate the web applications while working on the distributed real time, embedded and online system. These services are also called the application service which includes the combination of programming and data. Theses application services can be accessed through peer to peer networks and some other application services (like these).

Web Services is a kind of website which behaves in both the static and dynamic way, it not only provides the information in static way but also provide some dynamic way to users to interact with each other.

Before the concept of web services there is another technology initiated in 1995 named Distributed Common Object Model (DCOM), which allows common object model (COM) components to communicate across the local area networks and the wide area network. This technology uses the RPC mechanism send and receives the data between the COM components. By using DCOM technology a developer can instantiate an object on a remote server and by calling that results can be obtained. So than it moved towards CORBA and then after some research and development Web Services come in to exist.

## 7.1 Web Service Architecture and Workflow

The basic work flow of the Web services provide a simple way of performing remote procedure calls (RPC) over hyper text transfer protocol (HTTP) by using Simple Object Access Protocol SOAP. (SOAP actually a XML based technology, and this technology provides the way to call the procedure call over the HTTP, there is a SOAP remote server which is capable of understanding the procedure calls). Web services are completely described by its own particular language called Web Service Description Language (WSDL), this language applies the dynamic behaviour to web services at run time by providing the message formats, data types and protocols. XML schemas are also used by WSDL to provide the description. Web services don't require any kind of browser to run; these can be accessed from any plate form (Operating System) and the main reason of the platform independency is SOAP.

Web services have been categorised in to two major categories, explained below *[2]*

1. Exposing Web Services:
Exposing web services means to first make web services and the provide these to web servers to utilize these. When web services are exposed any web application can use these services by using the exposed methods.

2. Consuming Web Services:
Consuming services is the process to use the services at client side. This can be done through proxy, because through proxy all the call goes to the web server, looks like web service. There is a main reason to use the proxy; there is a lots communication over head on the HTTP, but we are shielded form details, so in this way WSDL is used to implement proxy with description of Web services.

To have successful interaction between the software some procedures are followed. Like there must be a connection, common communication protocol, so that messages can easily be exchanged. This behaviour of exchanging services between the requester and the provider is known as the semantics of web services.

Now agents (Agent: is software which is responsible for performing the autonomous behaviour is certain environment) have also been involved in the web services. To make web services more autonomous, intentional communication agents are involved.

This research now also has been enhanced from simple agent web services to multi agent web services where multiple agents are present to solve a common problem to achieve a common goal by acting in their own place and domain. For the multi agent communication a Foundation of Intelligent Physical Agents (FIPA) protocols are used. [3]

Web services are requested by some one from some provider and the whole procedure of communication between the requester and the provider is divided in to four steps as shown in First of all both requester and the provider must have to initiate to each other by their IDs.

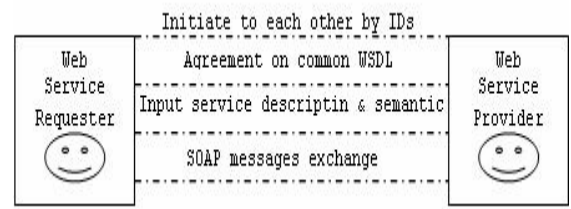

figure Web Service Request

1. Secondly, to have good communication a common compatible description language is required; both should be agreed on a WSDL & Semantic.
2. Thirdly service description and the semantics should be input via a hard coded agent in a general way.
3. SOAP messages are exchanged by both the requester as well as the provider agent.

## 7.2 Web Service Security (WSS)

Today the security is one of the hottest issues in the world of computer science especially regarding to the online application. Always people go for the best security mechanisms against the broker of security mechanisms; this is a non enable fight between makers and broker. So security is also required for the web services regarding to make web service secure, some kind of mechanisms are required for proper authentication, access to controls according to the defined roles, policies to secure distributed applications, data integrity, data confidentiality of messages, auditing and secure messaging. There are some mechanisms available like

- Some how SOAP is used to manage the securities, process of securing the web services via SOAP is happened in a way that request from the requester generated which has been taken by the SOAP which provides authentication, protection and confidentiality to message.

- WSS are also used token mechanism based on the cryptographic secure communication by involving some good and strong algorithms like Kerberos, Key management for encryption, authorizing, signing and verification.

- Security Assertion Markup Language mechanism, to allow user to login only once. This mechanism is composed of three components Assertion (used for the authentication purpose), Protocol (how a request receives and send) and Binding (used for the SOAP message exchange).

- XACML: this is a XML based mechanism for the writing the access controls rules and policies for different kinds of applications.

- Identity Web Services Framework (ID-WSF), this provides a mechanism to create, discover and update the identities through web services. [4]

## 7.3 Web Semantics

This is the extension of the Gird having all the services in proper well defined manners, and provide good environment to connect and work in interactive environment. Before the concept of web semantics web services require some kind of human involvement, but after the implementation of web semantic in to the web services, web services become automated in providing the composition and monitoring of the web services by providing the machine-interpretable description of the services. Darpa Agent Markup Language (DAM) is the language used for the development of the ontologies of the web semantics.

Web Services are becoming knowledge based; the knowledge attached with a service is of three kinds. First of all it is Service Profile these are basically the requirement and information required for the services. Second kind is Service Models these describes the way of working of the service and third is Service grounding which describes the details how an agent can access the service and how will the communication  will perform. [9]

## 8. WEB (GRID) SERVICES

As we have come now the formal details, architecture, work flow and the security issues of both the Web Services and Gird Computing. More over we have also been come to know the descriptive details about the Grid Services, Open Grid Service Infrastructure and Open Grid Service Architecture.

The main need of the time is to have Web Services having all its own functionalities as well as the features of the Grid computing. There are lots of similarities between these two different things and how can these both come join together to have new combination, we have already seen in the Grid Services and the OSGI & OSGA.

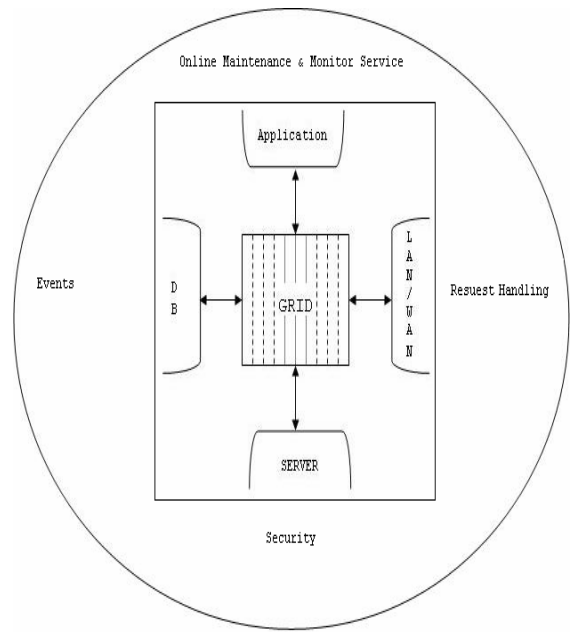

Figure Web (Grid) Service

The combination of these gives a distributed online technology, which includes all the behaviours of Grid Computing (Grid Service, OSGI & OSGA) in web services, is can be a new technology Web (Grid) Service as shown in Figure Web (Grid) Service. All the services provided by the Web Services are included as well as the complete functionalities of Grid computing based on the Open Grid Service Architecture.

Web (Grid) Service will also able to implement the concept of the virtual environment in the distributed online work environment between the heterogeneous entities. This will help in enhancing the concept of distributive online environment by enabling the technologies to work in a way by distributing and sharing the resources to have a good optimistic solution of the complex problem with in short time and cost.

The main architecture of Web (Grid) Service will be based on the Grid Computing and the Web Services, it will implements all the three layered architecture (when behaving in distributed environment) as well as provides the simple remote procedure call over HTTP for the communication, it will also prefers SOAP for this communication purposes. This will be described by using Grid Web Service Description Language (GWSDL). More over this will also be the platform independent.
Web (Grid) Service will also be able to provide to the ubiquitous access to systems, application with the implementation of Grid Computing, Integration of Grid Services and Web services.

Web (Grid) Service will also be capable of handling the security issues with more improvement in the Grid Computing and Web Services Security (WSS) architectures. Some possible basic securities like proper user authentication based on single sign on and cryptography, access to controls according to the defined roles, policies to secure distributed applications, data integrity, data confidentiality of messages, auditing and secure messaging. On the whole a combination of both the web services and grid computing technologies securities qualities can leads this new proposed technology to more new secure environment.

By introducing agents to this architecture, an autonomous behaviour can also be applied to Web (Grid) Services, where agent will be involved to perform the particular operation in a autonomous ways by standing on its own place following its allocated behaviour. And this can lead up to then Multi Agent System (MAS) to have Web (Grid) Autonomous Services based on the multiple agents responsible for the completion of a particular task.

Note: This is just proposed and discussed but requires more time to go in details and some proper implantation to have the experimental results, on the basis of those this can be approve as good technology or may requires more research to mature.

## 9. CHAIN WEB (GRID) SERVICES

The major requirement of the time with respect to the Web services and the Grid computing is to have a Chain Web (Grid) Services, this is a chain of web services, and the architecture of these web services is consists of both the web service contents and the grid (discussed before and shown also in the figure web (grid) service).
Many Web (Grid) Service will work in a network of chain, share the resources with each other in a fast and secure way, for the completion of particular task or to solve a problem in a joint way by making a distributed web (grid) service network, as shown in the figure Chain Web (Grid) Service.

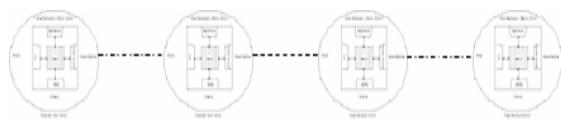

Figure Chain Web (Grid) Services

The most important issue related to Chain Web (Grid) Service which requires some time to solve and be maintained through out the time is the Security issue. With out having a strong security mechanism now days, it is not possible to have proper communication, interaction, sharing of resources and on time solution. Security requires in both the provider and requester's application and services. The major aim of this technology is to improve the sharing of resources in a fast and secure way with in virtual environments by ubiquitous having access for the solution of a particular problem.   So for this issue the security mechanism have been implemented which have been discussed in the Web (Grid) Services.

The multi agent system concept can also be applied to the chain web (grid) services to provide the autonomous behaviour.

Note: This is just proposed and discussed but requires more time to go in details and some proper implantation to have the experimental results, on the basis of those this can be approve as good technology or may requires more research to mature.

## 10. CONCLUSION

This report has been concluded with, Middleware technologies have been described in a good descriptive way, that what is actually middleware technology is, for what purposes it is beneficial, what are major and minor components of this, how much research in this field has been done before, what are the recent middleware technologies and in which areas still research is required and going on.

The brief introduction, architecture, work flow and the security issues of Grid Computing and the Web Services have also been described in the report. More over in this report Grid Service(s), Open Grid Service Infrastructure and Open Grid Service Architecture have been discussed to have a detail over view of the middle ware technologies.

Web (Grid) Service, a technology as the combination of both the Grid and the Web Service Technologies to become a single middle ware approach toward the application area related to the online as well as the distributed. This also handle the resource sharing, fast and quirk input and output of responses, decision making and secure commutation between the applications to share the data.

After having the discussion on the work flows, basic entities, security issues, basic requirement of the good architecture and need of the future of the middleware technologies, this report concluded with the an other middle ware approach Web (Grid) Service and the Chain Web (Grid) Services. Chain Web (Grid) Technology is a single middleware technology in the form of chain as a combination of many web services by holding the concept of Grid computing inside each web service and making them in the form of Web (Grid) Service to achieve a solution of the problem in an distributed online environment in a fast, quick and secure way by having ubiquitous access to the system.

In the end of the report, I (author) have been propose a new concept by merging the already existing concepts of Grid Computing and Web Services as Web (Grid) Services and then elaborate this concept in to the chain of Web (Grid) Services. This new concept requires some more time to make it into a proper shape. More over it also requires implementations, to have some experimental results, on the basis of these experimental results this can be finalised.

## 12. ACKNOWLEDGEMENT

This report has been written by me (Author: Zeeshan Ahmed) as the final report of Connectivity Software Technologies: Department of Intelligent Software Systems and Computer Science, Blekinge Institute of Technologies.

I am really thankful to my supervisor Dr. Prof. Rune, because of his guidance I was able to write and perform the research in the field of Middleware Technologies.

I am also thankful to Prof. Martin for his guidance in writing a report in very good and proper manners.